\documentclass[preprint,aps,showpacs,superscriptaddress,nofootinbib]{revtex4}
\usepackage[intlimits]{amsmath}
\usepackage{amssymb}
\usepackage{graphicx}
\usepackage{booktabs}
\usepackage{mathrsfs,slashed}
\usepackage{hyperref}
\usepackage{epsfig}
\usepackage{color}

\newcommand{\bea}{\begin{eqnarray}}
\newcommand{\eea}{\end{eqnarray}}

\DeclareMathOperator{\Tr}{Tr}
\DeclareMathOperator{\tr}{tr}

\begin{document}
\title{Variational path-integral approach to back-reactions of composite mesons in the Nambu-Jona-Lasinio model
}

\author{D.~Blaschke}
\email{david.blaschke@ift.uni.wroc.pl}
\affiliation{Instytut Fizyki Teoretycznej,
    Uniwersytet Wroc{\l}awski,
    pl. M. Borna 9,
    50-204 Wroc{\l}aw, Poland}
\affiliation{Laboratory of Theoretical Physics,
    Joint Institute for Nuclear Research,
    Joliot-Curie str. 6,
    141980 Dubna, Russia}
\affiliation{National Research Nuclear University (MEPhI),
    Kashirskoe Sho{ss}e 31,
    115409 Moscow, Russia}

\author{D.~Ebert}
\email{debert@physik.hu-berlin.de}
\affiliation{   Institut f\"ur Physik,
    Humboldt Universit\"at zu Berlin,
    Newtonstra{\ss}e 15,
    12489 Berlin, Germany}

\date{\today}
\begin{abstract}
For the investigation of back-reactions of composite mesons in the NJL model, a variational path-integral treatment is formulated which yields an effective action $\mathscr{A}_{\rm eff}[D_{\sigma},D_{\pi}; S]$,
depending on the propagators $D_\sigma$, $D_\pi$ of $\sigma-$ and $\pi-$mesons and on the full quark propagator $S$.
The stationarity conditions $\delta \mathscr{A}_{\rm eff}/ \delta S = 0$,
$\delta \mathscr{A}_{\rm eff}/ \delta D_\sigma = 0$, $\delta \mathscr{A}_{\rm eff}/ \delta D_\pi = 0$, then lead
to coupled Schwinger-Dyson (SD) equations for the quark self-energy and the meson polarization functions.
These results  reproduce and extend results of the so-called "$\Phi-$derivable" approach and provide a functional formulation for diagrammatic resummations of $1/N_c-$corrections in the NJL model.
Finally, we perform a low-momentum estimate of the quark and meson loop contributions to the polarization function of the pion and on this basis discuss the Goldstone theorem.
\end{abstract}

\maketitle

\section{Introduction}
It is well known that the Nambu-Jona-Lasinio (NJL) model \cite{1}
provides us in the leading order $1/N_c-$approximation ($N_c$ is the number of colors) with a relatively complete picture of low-energy meson physics \cite{2,3}.
There have been also interesting attempts to consider next-to-leading-order $1/N_c$ corrections, associated to composite meson exchange, contributing to the dynamical quark mass, the quark condensate and the meson polarization functions \cite{4,5,6,7,8,9,10}.
In particular, the approaches of Refs.~\cite{6,10} are based on a selfconsistent treatment of the Schwinger-Dyson (SD) equations for the quark propagator and the pion and sigma meson polarization functions.
By avoiding all functional methods, the SD equations in these papers were obtained by using 
quark propagators in the Hartree (H) approximation and composite meson propagators in the form of infinite sums of quark-antiquark polarization diagrams, usually considered in the so-called random-phase approximation (RPA) of many-body theory.
With the perturbative quark propagator in the Hartree form
(discarding the momentum-\-dependent contribution to the quark self-energy) and RPA meson propagators, it could be shown that the extended NJL model satisfies the Goldstone theorem, the Goldberger-Treiman and the Gell-Mann-Oakes-Renner (GMOR) relations.
Analoguous results were obtained in Ref.~\cite{7} using the effective action formalism.

In the present paper, we apply path-integral methods combined with $1/N_c$ arguments in order to derive the selfconsistently coupled SD equations for quarks and the composite meson polarization functions by variational methods.
Our derived effective action $\mathscr{A}_{\rm eff}[D_{\sigma},D_{\pi}; S]$ depends on the meson and quark propagators  $D_{\sigma}$, $D_{\pi}$ and $S$, respectively.
It contains besides the leading order quark loop ("dumbbell") diagram of order $\mathscr{O}(N_c)$ 
(see Fig.~\ref{fig:1}a) a correction term of
order  $\mathscr{O}(1)$ given by a quark loop with meson propagator (see Fig.~\ref{fig:1}b) and next-order
$\mathscr{O}(1/N_c)$ diagrams in the form of a quark loop with two crossed meson propagators 
(Fig.~\ref{fig:1}c) as well as diagrams with two quark loops connected by three meson propagators, compare Figs.~\ref{fig:1}d-e.
The corresponding order of a given loop diagram easily follows from the fact that a meson propagator contributes a factor $1/N_c$, $D_M = \mathscr{O}(1/N_c)$, whereas the color trace for a closed quark loop contributes a factor $N_c$.
It is interesting to note that the stationarity conditions of the effective action 
$\mathscr{A}_{\rm eff}[D_\sigma, D_\pi; S]$:
$\delta \mathscr{A}_{\rm eff}/ \delta S = 0$, 
$\delta \mathscr{A}_{\rm eff}/ \delta D_\sigma = 0$, $\delta \mathscr{A}_{\rm eff}/ \delta D_\pi = 0$
lead to the SD equations for the quark self-energy $\Sigma_M$ associated with the meson exchange and 
to the polarization functions $\Pi_M$ of composite mesons $M=(\sigma,\pi)$.

It is worth mentioning that the considered variational path-integral approach reproduces and partly enlarges the results of the so-called "$\Phi-$derivable" approximation \cite{11,12,13,14,15}.
Obviously, it gives also a functional foundation of diagrammatically motivated SD equations.
In the second part of the paper, we calculate in low-momentum approximation the relevant quark and meson loop diagrams appearing in the quark and meson SD equations.
Throughout this work we use RPA-like meson propagators and full quark propagators
which include the momentum-dependent contribution to the quark self-energy 
(see the meson-exchange diagram of Fig. 2 below).
Note that contrary to other approaches, our calculations explicitly include the meson correction diagram of order  $\mathscr{O}(1/N_c)$ in a nonperturbative way.
On this basis the status of the Goldstone theorem is discussed.

\section{Model and method}
Let us consider the Nambu-Jona-Lasinio (NJL) model for quarks with $N_f=2$ flavor and $N_c=3$ color
degrees of freedom given by the Lagrangian \cite{2,3} (henceforth we shall keep $N_c$ arbitrary)
\bea
\mathscr{L}= \mathscr{L}_0 + \mathscr{L}_{\rm int}~,
\eea
where
\bea
\mathscr{L}_0 = \bar{q} \left(i\slashed\partial -m_0 \right) q
\eea
is the free part with $\slashed \partial = \gamma_\mu \partial^\mu$ and containing a small current quark mass $m_0=m_0^u=m_0^d$ which explicitly breaks the chiral symmetry.
The interaction part 
\bea
\label{Lint}
\mathscr{L}_{\rm int}=\frac{G}{N_c}\left[\left(\bar{q} q \right)^2 + \left( \bar{q}i\gamma_5 \vec{\tau} q \right)^2 \right]
\eea
is a local four-fermion interaction term of strength $G/N_c$ for scalar isoscalar and pseudoscalar isovector quark-antiquark channels, respectively.
Here the quark fields $q(x)=\{q_i^a(x)\}$
are taken as Dirac spinors forming a flavor $SU(2)_f$ doublet ($i=1,2$).
and color $SU(N_c)-$plet ($a=1,\dots,N_c$), and $\tau_i$ are the isospin Pauli matrices.
Flavor and color indices are summed over and are suppressed in the following.

A central quantity for studying the interplay of quark and composite meson degrees of freedom is the
partition function which in the path-integral approach takes the form
\bea
\label{Z}
\mathscr{Z} = \int \mathscr{D} q \mathscr{D} \bar{q} \exp\left[ i \int d^4x \mathscr{L}(q,\bar{q})\right]~,
\eea
where here and henceforth normalization factors are absorbed into the integration measure.

In order to perform the integration over quark fields in Eq.~(\ref{Z}), one bi-linearizes the four-quark interaction term in Eq.~(\ref{Lint}) using the Hubbard-Stratonovich transformation and introducing color-singlet composite meson fields $\sigma$ and $\vec{\pi}$.
This yields \cite{2,3}
\bea
\label{Z-HS}
\mathscr{Z} = \int \mathscr{D} q \mathscr{D} \bar{q} \mathscr{D} \sigma \mathscr{D} \vec{\pi}
\exp\left[ i  \widetilde{\mathscr{A}}(q,\bar{q},\sigma,\vec{\pi})\right]~,
\eea
where the corresponding action is given by
\bea
\label{A}
\widetilde{\mathscr{A}}(q,\bar{q},\sigma,\vec{\pi})=\int d^4x d^4y && \left\{
\bar{q}(x)\left[ \left(i\slashed \partial _x - m_0\right)\delta^{(4)}(x-y) - \phi(x,y)\right] q(y) \right.
\nonumber\\
&&\left.
- \frac{N_c}{4G} \left[ \sigma^2(x) + \vec{\pi}^2(x) \right]\delta^{(4)}(x-y)
\right\}~,
\eea
and we have introduced for brevity
\bea
\phi(x,y) = \left[\sigma(x) + i\gamma_5\vec{\tau}\vec{\pi} \right] \delta^{(4)}(x-y)~.
\eea
As is well-known, the four-quark interaction leads to a non-vanishing vacuum expectation (mean-field)
value of the scalar field $\sigma_{\rm mf}$.
Let us therefore perform in Eq.~(\ref{A}) the field shift
$\sigma(x)=\sigma_{\rm mf}+\sigma^\prime(x)$.
Moreover, we find it useful to subtract and add a quark self-energy term $\Sigma_M(x,y)$,
\bea
\label{SigmaM}
\Sigma_M(x,y)=\Sigma_\pi(x,y) + \Sigma_\sigma(x,y)
\eea
in Eq.~(\ref{A}), describing the backreaction of mesons ($\pi,\sigma$) on the quark propagation,
which will be determined by a self-consistent variational principle later.
From now on we will omit the prime on the $\sigma-$field.
We then have
\bea
\label{A-mf}
\widetilde{\mathscr{A}}(q,\bar{q},\sigma,\vec{\pi})&=&\int d^4x d^4y
\left\{ \bar{q}(x)\left[ i S^{-1}(x,y) + \Sigma_M(x,y) - \phi(x,y)\right] q(y) \right.
\nonumber\\
&-&\left. \frac{N_c}{4G} \left[ \sigma^2(x) + \vec{\pi}^2(x)
\right]\delta^{(4)}(x-y) - \frac{N_c}{4G} \left[ \sigma_{\rm mf}^2
+ 2\sigma_{\rm mf}\sigma(x) \right]\delta^{(4)}(x-y) \right\}.
\eea 
In the following, the compensating (added) term
$\Sigma_M(x,y)$ in the first line of  Eq.~(\ref{A-mf}) will be
treated together with the meson fields $\phi(x,y)$ as an
interaction. $S^{-1}(x,y)$ is the inverse of the full quark
propagator given by \footnote{Assuming translational invariance,
we indeed have $S(x,y)=S(x-y)$, $\Sigma_M(x,y)=\Sigma_M(x-y)$. In
Eq.~(\ref{S}) we have explicitly shown the color and flavor
indices which, for brevity, are later omitted.} 
\bea 
\label{S}
iS^{-1}_{ai,bj} (x,y) =  \left[ \left(i\slashed \partial _x - m_0\right)\delta^{(4)}(x-y) - \Sigma(x,y)\right]
\delta_{ab}\delta_{ij} 
\eea 
with $\Sigma(x,y)$ being the total
dynamical quark self-energy \bea \label{Sigma} \Sigma(x,y) =
\Sigma_H \delta^{(4)}(x-y) + \Sigma_M(x,y)~, \eea where
$\sigma_{\rm mf}\equiv \Sigma_H$ is the so-called Hartree
(tadpole) mass which will be determined later, compare
Eqs.~(\ref{SigmaH}), (\ref{Hm}).

Let us now perform the path integral over quark fields in Eq.~(\ref{Z-HS}) by using Eqs.~(\ref{A-mf}), (\ref{S}) and a short hand matrix notation for both discrete group and continuous space-time indices,
\bea
&&\int \mathscr{D} q \mathscr{D} \bar{q} \exp\left\{
i \int d^4x d^4y \bar{q}(x) \left[ iS^{-1}(x,y) + \Sigma_M(x,y) - \phi(x,y)\right]q(y) \right\}
\nonumber\\
&&\hspace{2cm}= \det \left[ iS^{-1} + \Sigma_M - \phi \right]~.
\eea
The determinant is further rewritten in  the usual way
\bea
\label{det-exp}
\det \left[ iS^{-1} + \Sigma_M - \phi \right] &=&
\exp\left\{\Tr \ln \left[ iS^{-1} + \Sigma_M - \phi \right] \right\}
\nonumber\\
&=&\exp\left\{\Tr \ln iS^{-1} - \sum_{n=1}^\infty \frac{1}{n}\Tr\left[iS\left(\Sigma_M-\phi\right) \right]^n \right\}.
\eea

Note that the power-law multiplication in the Taylor expansion term in Eq.~(\ref{det-exp}) has to be
understood as matrix multiplication including integrations, and the trace $\Tr$ runs over discrete color, flavor and Dirac indices and over continuous space-time indices,
$\Tr \dots = {\rm tr_c}{\rm tr_f}{\rm tr_D}\int d^4x \dots \equiv {\rm tr}~{\rm tr}_x$; ${\rm tr}_x=\int d^4x \dots $.
Obviously, the second term in the second line of Eq.~(\ref{det-exp}) describes a quark-loop expansion in the form of ring diagrams consisting of full quark propagators $S$ emitting vertices $i(\Sigma_M - \phi)$.
By construction, the expression in the second line of Eq.~(\ref{det-exp}) does not depend on $\Sigma_M$,
because cancellations of corresponding infinitely many terms occur.
In order to get for the effective action an expression which really depends on the mesonic quark self-energy contribution $\Sigma_M$ and thus allows for a variational, self-consistent determination by a Schwinger-Dyson (SD) equation, one must avoid an identical cancellation.
This can be achieved by 
modifying the vertex factors in the series expansion in Eq.~(\ref{det-exp}) in discarding all 2P-reducible ring diagrams with higher powers $(\Sigma_M)^n$, $n>1$, and keeping only the necessary linear term in 
$\Sigma_M$.
The infinite sum in Eq.~(\ref{det-exp}) is thus changed by using the vertex truncation (see also further comments in footnote 2)
\bea
\label{trunc}
\Tr \left[iS(\Sigma_M-\phi) \right]^n \to \Tr \left[iS(\Sigma_M \delta_{n1}-\phi) \right]^n~.
\eea
In the following, we are interested to determine $\Sigma_M$ just in subleading order 
$\Sigma_M={\mathscr{O}}(1/N_c)$.
Using Eqs.~(\ref{A-mf}),(\ref{det-exp}) and (\ref{trunc}), the remaining mesonic path integral in the partition function is then given by
\bea
\label{Z-mes}
\mathscr{Z} = \int \mathscr{D} \sigma \mathscr{D} \vec{\pi}
\exp\left[ i  \widetilde{\mathscr{A}}_{\rm eff}(S,\sigma,\vec{\pi})\right]~,
\eea
where the truncated effective meson action $\widetilde{\mathscr{A}}_{\rm eff}$
is written up to fourth order terms $\mathscr{O}(\phi^4)$ as
\bea
 \widetilde{\mathscr{A}}_{\rm eff}(S,\sigma,\vec{\pi}) &=&
 - i \Tr \ln i S^{-1} - \Tr S \Sigma_M + \Tr S \phi - \frac{i}{2}\Tr (S \phi)^2 - \frac{1}{3} \Tr (S \phi)^3
 + \frac{i}{4}\Tr (S \phi)^4 + \dots \nonumber\\
 && - \frac{N_c}{4G} \int d^4 x (\sigma^2 + \vec{\pi}^2) - \frac{N_c}{4G} \Sigma_H^2 \int d^4 x
 - \frac{N_c}{2G}\Sigma_H \int d^4 x \sigma(x)~.
\eea
As usual \cite{2}, the Hartree mass $\Sigma_H$ has to be determined by the stationarity condition
\bea
\label{gap}
\frac{\delta \widetilde{\mathscr{A}}_{\rm eff}}{\delta \sigma(x)}\bigg|_{\phi=0} =
\tr S - \frac{N_c}{2G}\Sigma_H =0~,
\eea
which, after taking the trace over color and flavor indices, takes the standard form of the gap equation
\bea
\label{SigmaH}
\Sigma_H = 2G N_f \tr_D S(x,x) = - \frac{2G}{N_c} \langle \bar{q} q \rangle~.
\eea
Note that the traces over the color and flavor unit matrices in the quark propagator Eq.~(\ref{S})
yield a factor $N_cN_f$, and $S$ denotes now in Eq.~(\ref{SigmaH}) the quark propagator without color and flavor indices.
For completeness, Eq.~(\ref{SigmaH}) shows also the relation between the Hartree mass $\Sigma_H$ and
the quark condensate $\langle \bar{q} q \rangle$.
Obviously, in the large $N_c$ limit one has $\Sigma_H=\mathscr{O}(1)$.
Contrary, as will be shown in the following, the meson exchange contributions $\Sigma_M$ in
Eq.~(\ref{Sigma}) turn out to be next-to-leading order corrections $\mathscr{O}(1/N_c)$.
It is worth mentioning that linear terms in $\sigma$ and $\vec{\pi}$ vanish in the effective action
$\widetilde{\mathscr{A}}_{\rm eff}$ because of Eq.~(\ref{gap}) and
$\Tr S[i\gamma_5 \vec{\tau}\cdot \vec{\pi}]=0$.

Finally, we have yet to perform the path integral over meson fields in Eq.~(\ref{Z-mes}).
In order to get also a variational principle for the determination of the meson polarization functions
(self-energies) $\Pi_M$ by a SD equation, it is again convenient to first add and subtract quadratic terms in mesonic fields containing  $\Pi_{\sigma,\vec{\pi}}$. We then have
\bea
\label{Z-new}
\mathscr{Z}&=&\exp\left\{i\left[-i \Tr \ln iS^{-1} - \Tr S\Sigma_M\right]\right\}\nonumber\\
&&\times \int \mathscr{D} \sigma \mathscr{D} \vec{\pi}
\exp\left\{i\left\{\frac{1}{2}\left(\sigma i D_\sigma^{-1} \sigma + \vec{\pi} i D_\pi^{-1} \vec{\pi}\right)
-\frac{1}{2}\left( \sigma \Pi_\sigma \sigma + \vec{\pi} \Pi_\pi \vec{\pi}\right)
\right. \right. \nonumber\\
&&\left. \left. - \frac{N_c}{4G} \Sigma_H^2 \int d^4 x
-\frac{i}{2}N_f N_c \sigma \tr_D [SS]\sigma
-\frac{i}{2}N_fN_c\vec{\pi}  \tr_D [i\gamma_5 Si\gamma_5 S] \vec{\pi}
\right.\right. \nonumber\\
&&\left.\left. 
-\frac{N_fN_c}{3}\tr_D[S\sigma S\sigma S\sigma] -\frac{N_fN_c}{3}\tr_D[S\sigma S i\gamma_5\vec{\pi}Si\gamma_5\vec{\pi}]
\right.\right.\nonumber\\
&&\left.\left. 
+\frac{iN_fN_c}{4}\tr_D[S\Gamma_{M_1}S\Gamma_{M_2}S\Gamma_{M_3}S\Gamma_{M_4}]
\left[\sigma^4 + 2 \sigma^2 \vec{\pi}^2 + \left(\vec{\pi}^2\right)^2\right] +\dots~, \right\} \right\} ~,
\eea
where for brevity, integration is included into multiplication and
$\Gamma_{M_i}=(1,i\gamma_5)$ are vertices suitably chosen for the corresponding powers of meson fields. 
Note that the inverse propagators of composite mesons $\sigma, \vec{\pi}$
are defined by
\bea 
\label{Ip} 
i{D}^{-1}_{ \sigma ,\pi}
\left(x,y\right)=\left(-\frac{N_c}{2G}\delta^4\left(x-y\right)+\Pi_{\sigma,\pi}\left(x,y\right)
\right), 
\eea
and the subtracted quadratic terms in Eq.~(\ref{Z-new}) containing $\Pi_{\sigma,\pi}$ are again treated
as perturbations. 
Using the cumulant expansion and considering only $1P$- and $2P$-irreducible connected path-integral averages, the path-integral in Eq.~(\ref{Z-new}) leads to the expression
\bea 
\label{Z-new2}
\mathscr{Z}\left[{D}_{\sigma,\pi};S\right]&=&\exp \left\{ i \mathscr{A}_{\rm eff}
\left[{D}_{\sigma ,\pi};S\right] \right\} \nonumber\\
&=& \exp \left\{ i \left\{-i\Tr \ln iS^{-1} - \Tr S\Sigma_M\ -\frac{N_c}{4G}\Sigma^2_H \int d^4 x 
\right.\right.
\nonumber\\
&&\left. \left. 
+ \frac{i}{2}\tr_x \ln i {D}^{-1}_{\sigma}
+\frac{3i}{2}\tr_x \ln i {D}^{-1}_\pi 
\right. \right.
\nonumber\\
&&\left. \left. 
-\frac{1}{2}\tr_x {D}_{\sigma}\Pi_{\sigma}
-\frac{3}{2}\tr_x {D}_{\pi}\Pi_{\pi}-\frac{i}{2} N_f N_c
\tr_{{D}}\left[SS\right]{D}_{\sigma}
\right. \right.
\nonumber\\
&&\left. \left. 
-\frac{i}{2} 3N_f N_c \tr_{{D}}\left[i\gamma_5 S i \gamma_5 S \right] {D}_{\pi}
\right. \right.
\nonumber\\
&&\left. \left. 
+3! \frac{i}{2}\left( \frac{N_f}{3} N_c\right)^2
\tr_{{D}} \left[S S S\right] {D}^3_{\sigma}
\tr_{{D}}\left[SSS\right]
\right. \right.
\nonumber\\
&&\left. \left. 
+2!\frac{i}{2}\left( \frac{N_f}{3} N_c\right)^2 3
\tr_{{D}} \left[S S i\gamma_5  S i \gamma_5\right] {D}_{\sigma}
{D}^2_{\pi} \tr_{{D}}\left[SSi\gamma_5  S i \gamma_5\right] 
\right. \right.
\nonumber\\
&& \left. \left. 
+\frac{i}{4}N_f N_c \tr_{{D}} \left[S S S S
\right] {D}^2_{\sigma} +\frac{3i}{4}N_f N_c
\tr_{{D}} \left[S S i\gamma_5  S S i \gamma_5\right] {D}_{\sigma} {D}_{\pi} 
\right. \right.
\nonumber\\
&& \left. \left. -3i\frac{N_f}{4} N_c \tr_{{D}} \left[S i\gamma_5
S i\gamma_5  S i\gamma_5 S i \gamma_5\right] {D}^2_{\pi}
\right\}\right\}~.
 \eea

Here the space-time integration in higher terms has again been included in
the matrix multiplication. 
Again we have considered only the first cumulants 
$\frac{1}{2}{D}_{\sigma}\Pi_{\sigma}$,
$\frac{3}{2}{D}_{\pi}\Pi_{\pi}$ of the interaction terms
$-\frac{1}{2}\sigma\Pi_{\sigma}\sigma$,
$-\frac{1}{2}\pi_i\Pi_{ij,\pi}\pi_j$ in order to avoid
cancellations with identical terms arising from the ring expansion
of $\frac{i}{2}\tr_x \ln i {D}^{-1}_{\sigma,\pi}$.

\begin{figure*}[!htb]
\includegraphics[height=0.3\textwidth]{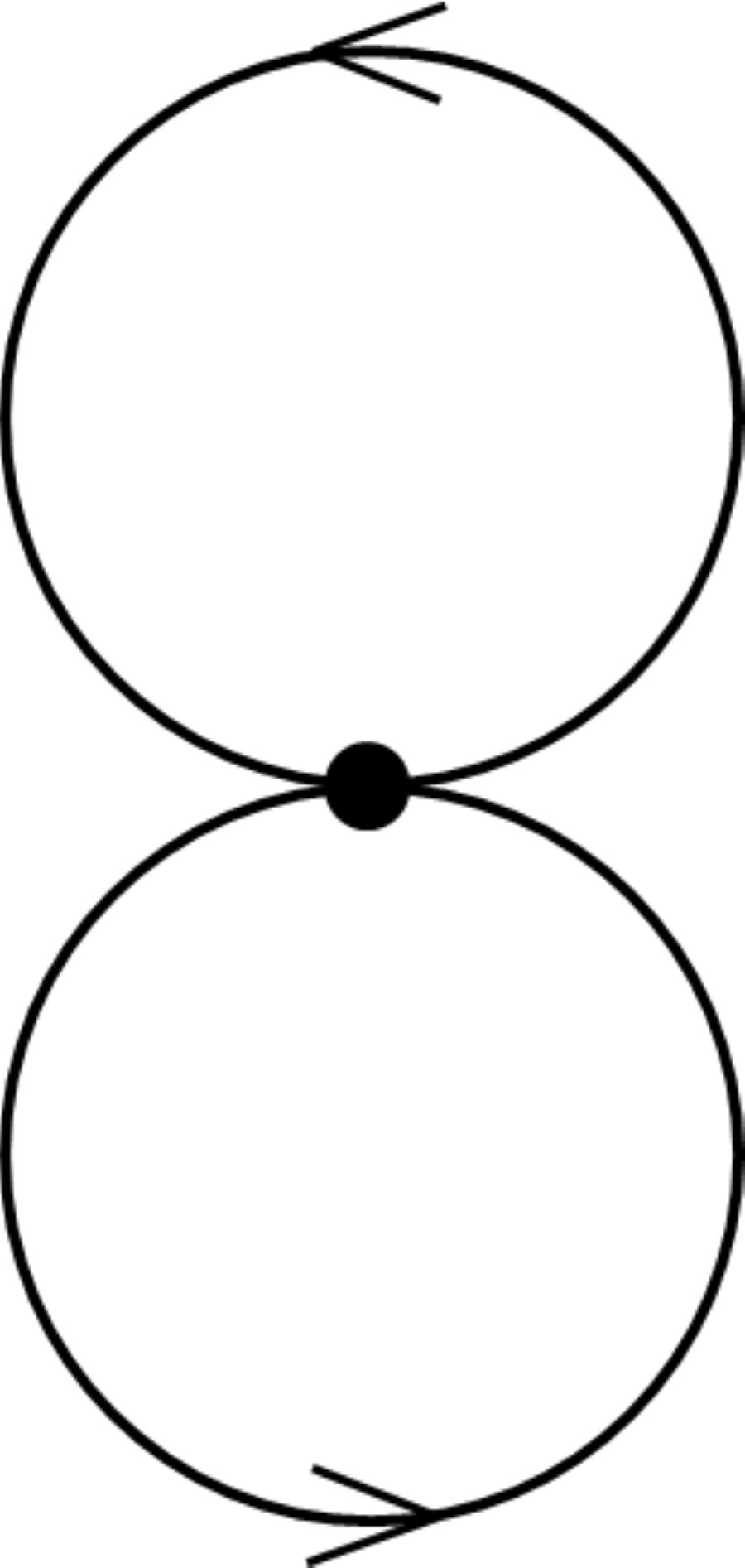}\hfill
\includegraphics[height=0.22\textwidth]{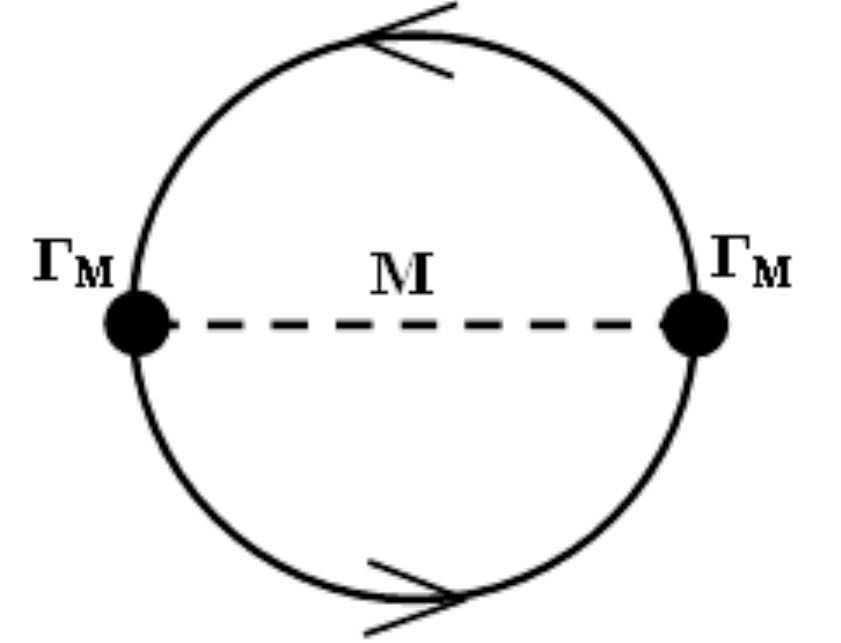}\hfill
\includegraphics[height=0.22\textwidth]{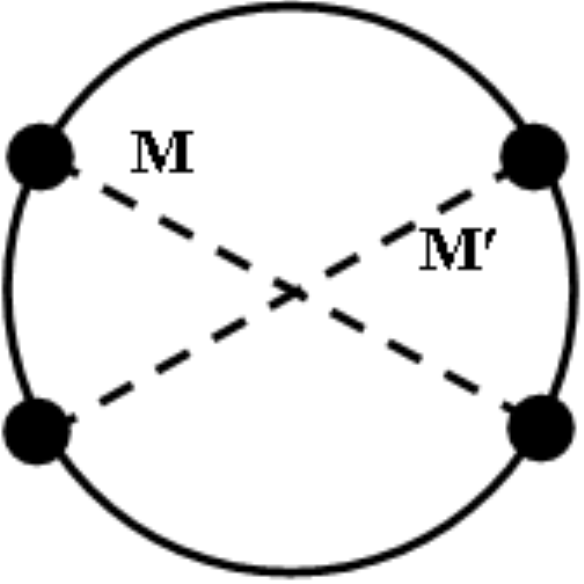}
\\
(a)\hspace{6cm} (b) \hspace{6cm} ({c})
\\[5mm]
\includegraphics[height=0.2\textwidth]{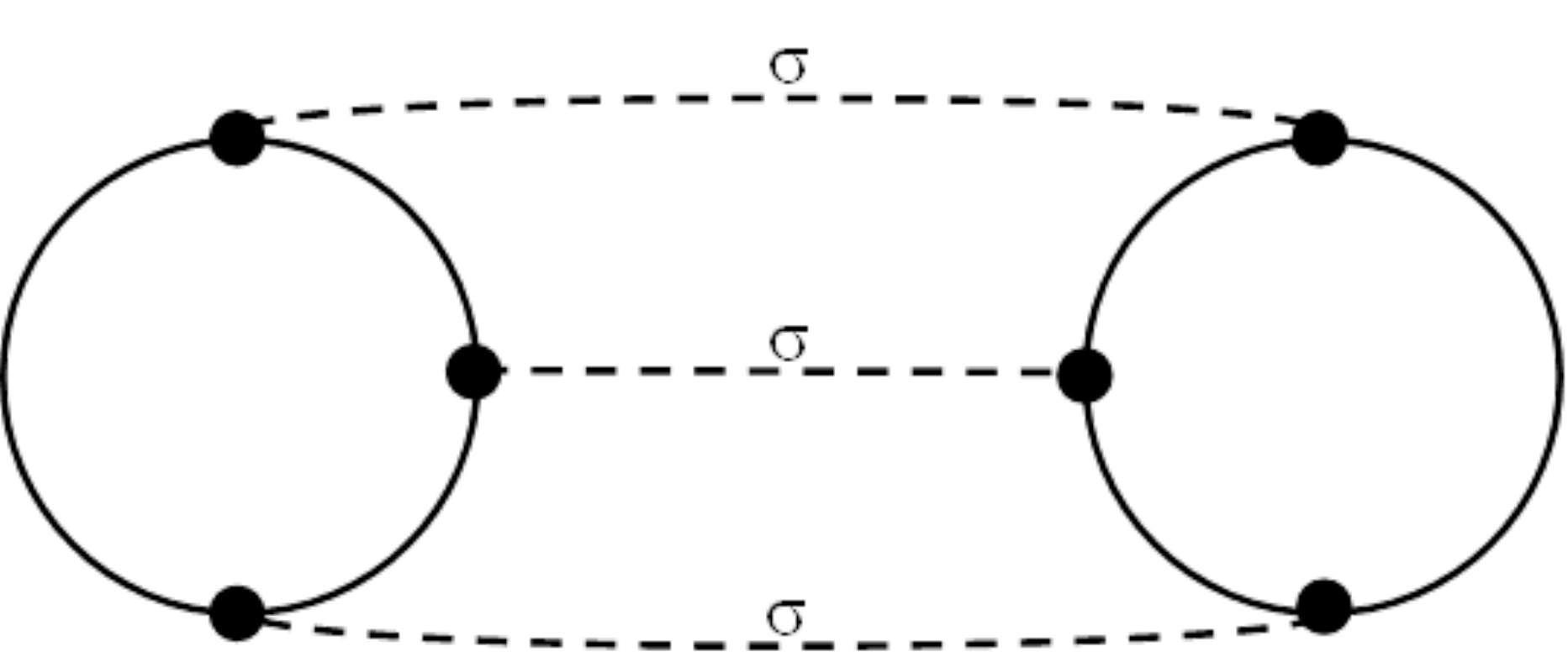}\hfill
\includegraphics[height=0.2\textwidth]{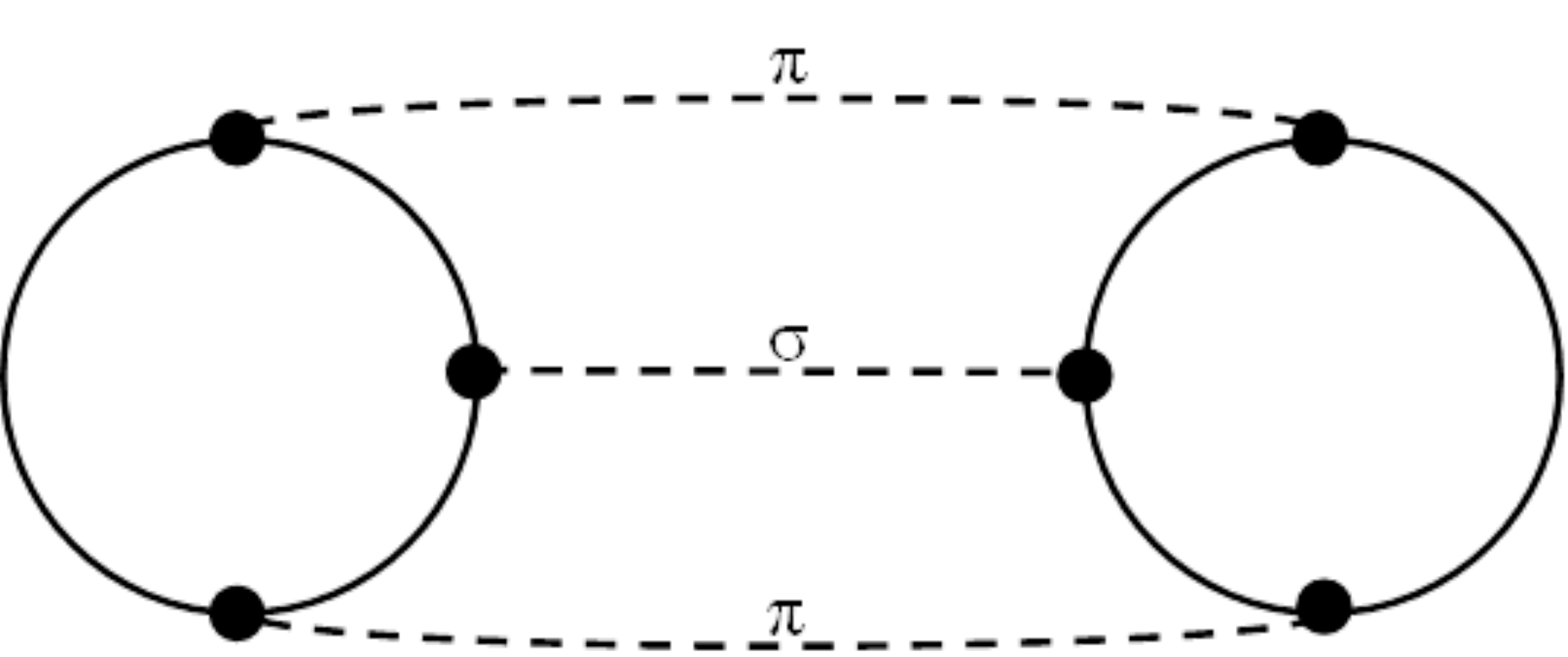}
\\
(d)\hspace{8cm} (e) 
\caption{2P-irreducible loop diagrams described by the effective
action $\mathscr{A}_{\rm eff}$ of Eq.~(\ref{Z-new2}). Solid lines
correspond to the full quark propagator $S$ of Eq.~(\ref{S}),
dashed lines denoted by $M=(\sigma,\pi)$
describe the RPA-like meson propagator
${D}_{\sigma,\pi}$ of Eq.~(\ref{Ip}), and
$\Gamma_M=\left(1,i\gamma_5\right)$. 
Taking into account that
${D}_{\sigma,\pi}=\mathscr{O}\left(1/N_c\right)$ and that every
quark loop contributes a color factor $N_c$, it follows that
the first diagram (a) is of order $\mathscr{O}(N_c)$, the next diagram 
(b) is of order $\mathscr{O}(1)$, whereas the diagrams ({c}) - (e) 
are of subleading order $\mathscr{O}\left(1/N_c\right)$.
\label{fig:1}}
\end{figure*}

Fig.~\ref{fig:1} shows typical loop diagrams with meson exchange described by
the effective action
$\mathscr{A}_{\rm eff}\left[{D}_{\sigma,\pi};S\right]$

\begin{figure*}[!htb]
\includegraphics[width=0.7\textwidth]{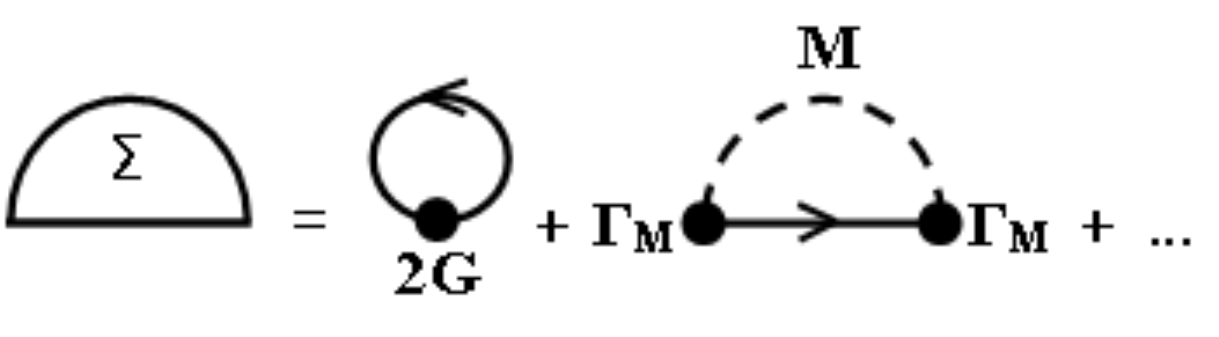}
\caption{Graphical representation of the SD-equation for the quark
self-energy (compare with Eqs.~(\ref{Sigma}), (\ref{SigmaH}) and Eq.~(\ref{E})). 
The first term in r.h.s. represents the Hartree term
$\Sigma_H=\mathscr{O}(1)$, the second diagram shows the
$\mathscr{O}\left(1/N_c\right)$ contribution $\Sigma_M$ arising from
$(\pi,\sigma)$-meson exchange; $\Gamma_M$ denotes the
meson-quark vertex factors (summation over $M=(\sigma,\pi)$ is understood). 
\label{fig:2a}
}
\end{figure*}

In order to determine the meson-exchange contribution $\Sigma_M$
to the quark self-energy and the meson polarization function
$\Pi_M$, we shall now require stationarity of the effective action
$\mathscr{A}_{\rm eff}\left[{D}_{\sigma,\pi};S\right]$ under
variations $\delta S_{ab}$ and $\delta {D}_{\sigma,\pi}$ of
quark and meson propagators. It is then easily seen that the
stationarity condition ${\delta\mathscr{A}_{\rm eff}}/{\delta S_{ab}}=0$ 
leads to a SD-equation for the meson-exchange contribution\footnote{
Note that the variation of $\Sigma_H^2 \sim (\tr_D S)^2$ in the third term 
of the second line in Eq.~(\ref{Z-new2}) gets cancelled by the term 
$-\Tr \left(S\, \delta \Sigma_M \right)$.
This can be seen by using (in shorthand notation) 
$\Sigma_M=\Sigma-\Sigma_H=iS_H^{-1} - i S^{-1}$,
where $S_H$ is the Hartree-type quark propagator
with the mean-field (Hartree) mass $~m_0+\Sigma_H$.
Note further that including, e.g., the quadratic ($n=2$) term of the series 
in Eq.~(\ref{det-exp}), its variational contribution would cancel the required linear term 
 $-\Tr \left(\delta S\, \Sigma_M \right)$ and thus would spoil the derivation of Eq.~(\ref{E}).
 Obviously, such unwanted cancellation is just avoided by the truncation prescription of Eq.~(\ref{trunc}). 
}
\bea 
\label{E}
\Sigma_M\left(x-y\right)&=&\Sigma_{\pi}\left(x-y\right)+\Sigma_{\sigma}\left(x-y\right)\nonumber\\
&=& - 3i \left[i\gamma_5 S(x-y)i\gamma_5\right] {D}_{\pi}(x-y) 
-i S(x-y) {D}_{\sigma}(x-y) + \dots ~,
\eea
which is graphically shown as the second term on the r.h.s. of Fig.~\ref{fig:2a}. 
Note that $\Sigma_M=\mathscr{O}\left(1/N_c\right)$. 
The ellipsis ... in Eq.~(\ref{E}) denote higher order self-energy
contributions arising from applying the functional derivative 
${\delta}/{\delta S_{ab}}$ to diagrams Figs.~\ref{fig:1} ({c}) - (e). 
Restricting us here to the subleading order term $\Sigma_M=\mathscr{O}\left(1/N_c\right)$, they will
be discarded in the following.

\begin{figure*}[!htb]
\includegraphics[width=\textwidth]{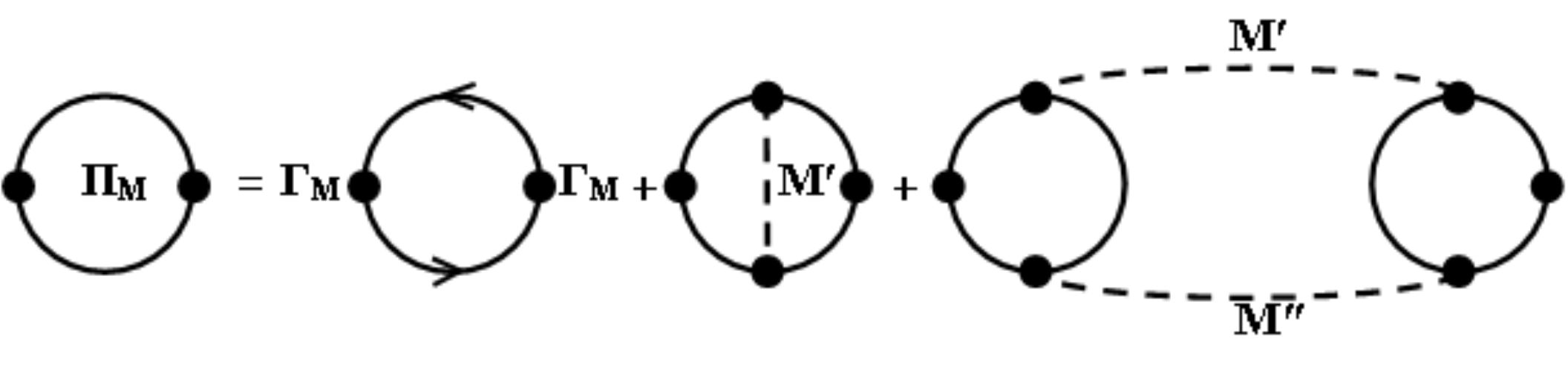}
\caption{Graphical representation of the SD-equations for the 
meson polarization function $\Pi_M$.  
The figure shows relevant leading and subleading contributions to the 
meson polarization function: 
the first term is the leading $\mathscr{O}(N_c)$ contribution from
the quark loop, whereas the second and third terms are
subleading $\mathscr{O}(1)$-contributions arising from a
correction to the meson-quark vertex and from two-meson
intermediate states, respectively. 
In the latter case the white circles are
related to the quark-loop vertices of Fig.~\ref{fig:1} (d) and (e)
(summation over meson propagators denoted shortly by $M^\prime$ and 
$(M^\prime,M^{\prime\prime})$ is understood).
\label{fig:2b}
}
\end{figure*}

Analogously, the stationarity conditions
${\delta\mathscr{A}_{\rm eff}}{\delta {D}_{\sigma,\pi}}=0$
determine the polarization functions $\Pi_{\sigma,\pi}$ of the
$\sigma$ and $\pi$ mesons self-consistently by the SD equation,
graphically shown in Fig.~\ref{fig:2b}. 
For example, we obtain
\bea 
\label{Aeff}
\frac{\delta\mathscr{A}_{\rm eff}}{\delta {D}_{\sigma}}&=& 
-\frac{1}{2}\Pi_{\sigma}-\frac{i}{2}N_f N_c \tr_{{D}}\left[SS\right] 
+\frac{i N_f N_c}{4} 2 \tr_{{D}}\left[SSSS\right] {D}_{\sigma} \nonumber\\
&&+ i \frac{3}{4} N_f N_c \tr_{{D}}\left[S S i\gamma_5 S S
i \gamma_5\right] {D}_{\pi} 
+3!\frac{i}{2}\left(\frac{N_f}{3}N_c\right)^2 \cdot 3
\tr_{{D}}\left[S S S\right] {D}^2_{\sigma}
\tr_{{D}}\left[S S S\right] \nonumber\\
&&+2!\frac{i}{2}\left(\frac{N_f}{3}N_c\right)^2 \cdot 3
\tr_{{D}}\left[S S i \gamma_5 S i \gamma_5\right]
{D}^2_{\pi} \tr_{{D}}\left[S S i \gamma_5 S i \gamma_5\right]=0.
 \eea

In an analogous way, one obtains the SD equation for $\Pi_{\pi}$.
It is worth emphasizing that Eqs.~(\ref{Z-new2}), (\ref{E}),
(\ref{Aeff}) and Figs.~\ref{fig:1}, \ref{fig:2a} and \ref{fig:2b}  of the present variational
path-integral approach reproduce (see Fig.~\ref{fig:1} (b)), as well as extend
(see Fig.~\ref{fig:1} ({c}) - (e)) results of the $\Phi$-derivable approach \cite{11,12,13,14,15}. 
Note also that our variational path-integral method provides a functional foundation 
for the diagrammatic treatment of $1/N_c$ corrections to the NJL model considered in
Refs. \cite{5,6,10}.

\section{Renormalized quark and meson propagators and Schwinger-Dyson equations}
\subsection{Hartree gap equation\label{ssec:hartree}}

By definition, quarks are not confined in the NJL-model and the
total dynamical quark mass $\tilde{m}$ is given by the pole of the
full quark propagator. In the pole approximation (see subsection
\ref{ssec:q-prop}), the full quark propagator takes the form
\bea 
\label{SS} S(x-y)&=&\int \frac{d^4p}{(2\pi)^4} e^{-ip(x-y)}S({p})~, \nonumber\\
S({p})&=&\frac{i Z}{\slashed{p}-\tilde{m}}~, 
\eea
where $Z$ is a renormalization constant. 
The gap equation (\ref{gap}) for the Hartree mass $\Sigma_H$ is then rewritten as
\bea 
\label{Hm} 
\Sigma_H(\tilde{m})=i 8 N_f Z G \tilde{m} I_1(\tilde{m})~,
\eea
where the momentum integral
\bea 
\label{Im} 
I_1(\tilde{m})=\int\frac{d^4k}{(2\pi)^4} \frac{1}{k^2-\tilde{m}^2}~,
\eea
has to be suitably regularized.

\subsection{Renormalized full quark propagator and total quark mass $\tilde{m}$
\label{ssec:q-prop}}
Let us next determine the quark self-energy $\Sigma(p,\tilde{m})$ depicted in Fig.~\ref{fig:2a}
and considered in momentum space
\bea 
\label{qse}
\Sigma(p,\tilde{m})=\Sigma_H(\tilde{m})+\Sigma_{\pi}(p,\tilde{m})+\Sigma_{\sigma}(p,\tilde{m})~.
\eea
For this aim, we need the Fourier-transformed expression of the meson propagator Eq.~(\ref{Ip}) 
which we conveniently rewrite in terms of the renormalized coupling constant $G_r$ and renormalized
polarization function $\Pi_M^r(k)$, $M=(\sigma,\pi)$. 
Using
\bea 
\label{GP} 
G_r=Z^2G, ~ \Pi_M^r(k)=Z^{-2}\Pi_M(k),
 \eea
we have 
\bea 
\label{DMk} 
{D}_M(k)=i Z^{-2}\left[-\frac{N_c}{2G_r}+\Pi^r_M(k)\right]^{-1} 
\eea 
and applying the pole approximation (see subsection \ref{ssec:RMP}), 
\bea
\label{DMk-ap} 
{D}_M(k)=i Z^{-2}\frac{g_{Mqq}^2}{k^2-m_M^2}. 
\eea
In expression (\ref{DMk-ap}) $g^2_{Mqq}$ is the square of the
induced meson-quark coupling constant,
\bea 
\label{g2} 
g^2_{Mqq}=\left(\frac{\partial \Pi^r_M}{\partial k^2}\right)^{-1}\bigg|_{k^2=m_M^2}~,
\eea 
and $m_M$ is the meson mass. 
For large $N_c$, the dominant quark loop contribution in
Fig.~\ref{fig:2b} leads to $\Pi_M^r=\mathscr{O}(N_c)$ so that
$g^2_{Mqq}=\mathscr{O}\left(1/N_c\right)$. 
Using Eqs.~(\ref{E}), (\ref{SS}), and (\ref{DMk-ap}), we obtain for the pion contribution
\bea 
\label{Epion}
\Sigma_{\pi}\left(p,\tilde{m}\right)&=&-3i\int\frac{d^4k}{(2\pi)^4}i\gamma_5S(k)i\gamma_5 {D}_{\pi}(p-k) \nonumber\\
&=& -3i g^2_{\pi qq}Z^{-1} \int \frac{d^4k}{(2\pi)^4}
\left(\frac{-\slashed{k}+\tilde{m}}{k^2-\tilde{m}^2}\right)\frac{1}{(p-k)^2-m_{\pi}^2}~,
 \eea
where the factor $3$ occurs from the isospin multiplicity of the pion.
Analogously,
\bea 
\label{Esigma}
\Sigma_{\sigma}\left(p,\tilde{m}\right)&=&
-i\int\frac{d^4k}{(2\pi)^4}S(k) {D}_{\sigma}(p-k) \nonumber\\
&=& i g^2_{\sigma qq}Z^{-1} \int
\frac{d^4k}{(2\pi)^4}\left(\frac{\slashed{k}+\tilde{m}}{k^2-\tilde{m}^2}\right)\frac{1}{(p-k)^2-m_{\sigma}^2}~.
\eea
Obviously, the momentum integrals (\ref{Epion}), (\ref{Esigma})
require a regularization. 
It is now convenient to decompose the quark self-energy from meson-exchange into two terms, 
e.g.,
\bea 
\label{Etwo}
\Sigma_{\pi}\left(p,\tilde{m}\right)=\Sigma_{\pi,S}\left(p^2,\tilde{m}\right)+\slashed{p}
\Sigma_{\pi,V}\left(p^2,\tilde{m}\right)
 \eea
containing two scalar functions $\Sigma_{\pi,S}$ and $\Sigma_{\pi,V}$. 
They are determined by
\bea 
\label{Es}
\Sigma_{\pi,S}\left(p^2,\tilde{m}\right)=\frac{1}{4}\tr_{{D}}\Sigma_{\pi}\left(p,\tilde{m}\right)~,
 \eea
\bea \label{Ev}
\Sigma_{\pi,V}\left(p^2,\tilde{m}\right)=\frac{1}{4p^2}\tr_{{D}}\left[\slashed{p}\Sigma_{\pi}\left(p,\tilde{m}\right)\right]~,
 \eea
and given explicitly as
\bea 
\label{Es2} 
\Sigma_{\pi,S}\left(p^2,\tilde{m}\right)= -3i g^2_{\pi qq}Z^{-1} \tilde{m}I_2(p^2;\tilde{m},m_\pi)~,
 \eea
\bea 
\label{Ev2}
\Sigma_{\pi,V}\left(p^2,\tilde{m}\right)= - \frac{1}{p^2}3i g^2_{\pi qq}Z^{-1} I_3(p^2;\tilde{m},m_\pi)~,
\eea
where $I_2$, $I_3$ are integral expressions given by
\bea 
\label{I2} 
I_2\left(p^2;\tilde{m},m_M\right)= \int\frac{d^4k}{(2\pi)^4}\frac{1}{k^2-\tilde{m}^2}\frac{1}{(p-k)^2-m_{M}^2}~,
\eea
\bea 
\label{I3} 
I_3\left(p^2;\tilde{m},m_M\right)=\int\frac{d^4k}{(2\pi)^4}\frac{-p\cdot k}{k^2-\tilde{m}^2}
\frac{1}{(p-k)^2-m_{M}^2},
\eea
with $m_M=(m_{\pi},m_\sigma)$.

Note that in the case of cutoff-regularizations, loop diagrams
containing meson propagators should be regularized with a momentum
cutoff $\tilde{\Lambda}$ differing from the cutoff $\Lambda$ used
in pure quark loop diagrams\footnote{This problem does not occur when using the formfactor 
regularization that is inherent in nonlocal generalizations of the NJL model. See, e.g., Ref.~\cite{8}.} 
\cite{6,9}.

Analogous expressions are obtained for $\Sigma_{\sigma,S}$ and $\Sigma_{\sigma,V}$. 
$\Sigma_{\sigma,S}$ takes the form of Eq.~(\ref{Es2}) with the replacements 
$-3g^2_{\pi qq} \rightarrow g^2_{\sigma qq}$, $m^2_{\pi}\rightarrow m^2_{\sigma}$, whereas
$\Sigma_{\sigma,V}$ takes the form of Eq.~(\ref{Ev2}) with the
replacements $3g^2_{\pi qq}\rightarrow g^2_{\sigma qq}$, $m^2_{\pi}\rightarrow m^2_{\sigma}$.

As supposed in Eq.~(\ref{SS}), the full dynamical quark
mass $\tilde{m}$ is given by the pole of the quark propagator. 
In order to determine $\tilde{m}$ in a self-consistent way, let us
write the Fourier transform of Eq.~(\ref{S}),
\bea 
\label{Sp} 
S(p)=i \left[\slashed{p}-m_0-\Sigma(p)\right]^{-1}
 \eea
in the form
\bea 
\label{Spz} 
S(p)=iZ(p^2) \left[\slashed{p}-\tilde{\Sigma}(p^2)\right]^{-1}
 \eea
with [5]
\bea 
\label{SigmaPM}
\tilde{\Sigma}\left(p^2,\tilde{m}\right)=Z(p^2)\left[m_0+\Sigma_H(\tilde{m})+\Sigma_{\pi,S}\left(p^2,\tilde{m}\right)+\Sigma_{\sigma,S}\left(p^2,\tilde{m}\right)\right]
 \eea
and $Z(p^2)$ being a running renormalization factor
\bea 
\label{Zp2}
Z(p^2)=\left[1-\Sigma_{\pi,V}(p^2,\tilde{m})-\Sigma_{\sigma,V}(p^2,\tilde{m})\right]^{-1}~.
 \eea

Obviously, the pole of the propagator Eq.~(\ref{Spz}) determines the full dynamical quark mass $\tilde{m}$ as \footnote{There is also a convenient off-shell renormalization at $p=0$,
$S(p)=i Z(0)\left[\slashed{p}-\tilde{\Sigma}(0)\right]^{-1}$ with $\tilde{m}(0)=\tilde{\Sigma}(0)$.}
\bea 
\label{P2} 
p^2= \tilde{\Sigma}^2(p^2,\tilde{m})|_{p^2=\tilde{m}^2}\equiv \tilde{m}^2~.
 \eea

Thus, in the pole approximation the full quark propagator (\ref{Spz}) indeed takes the form assumed in 
Eq.~(\ref{SS}) with the renormalization constant $Z=Z(p^2)|_{p^2=\tilde{m}^2}$. 
For completeness, let us quote a self-consistency gap equation for $\tilde{m}$ following from  Eqs.~(\ref{Sp}), (\ref{SigmaPM}) using  Eq.~(\ref{Es2}),
\bea 
\label{gapM} 
\tilde{m}=m_{0r}+\tilde{m}[8N_f i G_r I_1(\tilde{m}) &-& 3ig^2_{\pi qq}I_2(\tilde{m}^2;\tilde{m},m_{\pi})
+ i g^2_{\sigma qq}I_2(\tilde{m}^2;\tilde{m},m_{\sigma}]
 \eea
with $m_{0r}=Z m_0$ and $G_r=Z^2G$ being a renormalized current
quark mass and four-quark coupling constant, respectively.

Note that the expression Eq.~(\ref{gapM}), calculated with the full quark mass $\tilde{m}$ formally agrees with the result of Ref.~\cite{5} calculated with the mean field (Hartree) mass $m=m_0+\Sigma_H$.
Up to this difference, the $\pi-$ and $\sigma-$ corrections to the quark self-energy are again opposite in sign and cancel each other in part.
In particular, it is seen that the $\pi-$ correction is dominating due to its higher isospin degeneracy factor 3.   

Finally, using Eq.~(\ref{Ev2}) as well as the analogous expression for $\Sigma_{\sigma,V}$ and 
Eq.~(\ref{Spz}), $Z=Z(p^2=\tilde{m}^2)$ can be written as
\bea 
\label{Z-new3} Z=1 &-& \frac{1}{\tilde{m}^2} 3ig^2_{\pi qq}I_3(\tilde{m}^2;\tilde{m},m_{\pi})
-\frac{1}{\tilde{m}^2} i g^2_{\sigma qq}I_3(\tilde{m}^2;\tilde{m},m_{\sigma}).
 \eea

Note that since $g^2_{M qq} = \mathscr{O}(1/N_c)$, the meson exchange contributions to the full dynamical quark mass $\tilde{m}$ and the Z-factor are obviously next-to-leading order
contributions $\mathscr{O}(1/N_c)$.

\subsection{Renormalized meson propagators and polarization functions; composite meson masses
\label{ssec:RMP}}

\subsubsection{Meson propagators \label{sssec:MP}}

Let us next comment on the pole approximation for the meson
propagators ${D}_{\sigma,\pi}(k)$ quoted in Eq.~(\ref{DMk-ap}) and used in the calculation of the quark
self-energies. 
Clearly, the composite meson masses are determined by the poles of the meson propagators, i.e. by the vanishing of their denominators shown in Eq.~(\ref{DMk})
\bea 
\label{MMv} 
-\frac{N_c}{2G_r}+\Pi_M^r(m_M^2)=0.
 \eea
By subtracting the expression Eq.~(\ref{MMv}) in the denominator of ${D}_M(k)$, expanding around $k^2=m_M^2$ and retaining only the leading term, one obtains the expression
\bea 
\label{iD-1} 
i{D}^{-1}_M(k)&=& Z^2\left[\Pi_M^r(k)-\Pi_M^r(m_M^2)\right]\nonumber\\
&\approx& (k^2-m_M^2)Z^2
\frac{\partial\Pi_M^r}{\partial k^2}\bigg|_{k^2=m_M^2}\nonumber\\
&=& (k^2-m_M^2)Z^2g^{-2}_{Mqq}.
 \eea
Obviously, this is just the meson propagator in the pole approximation, Eq.~(\ref{DMk-ap}), with $g^2_{Mqq}$ given by Eq.~(\ref{g2}).\footnote{The expression (\ref{iD-1}) contains the squared 
coupling constant $g^2_{Mqq}$ and presents more precisely the $T$-matrix expression and not a pure propagator. Nevertheless, for brevity, it is called propagator.}

\subsubsection{Quark-loop contribution to polarization functions 
\label{sssec:Q-loop}}

The polarization function $\Pi_{\sigma}(k)$ is given by the SD-equation Eq.~(\ref{Aeff}) (with an analogous expression for $\Pi_{\pi}(k)$), shown graphically in Fig.~\ref{fig:2b}. 
Let us first consider the leading $\mathscr{O}(N_c)$ contribution from the quark loop diagram given by the first term on the r.h.s. of Fig.~\ref{fig:2b}. 
Using the pole approximation (\ref{SS}) for the quark propagator, one has
\bea 
\label{Psk} 
\Pi_{\sigma}(k)&=&-i N_f N_c\tr_{{D}}\int \frac{d^4p}{(2\pi)^4}S(p)S(p-k) \nonumber\\
&=& Z^2\left\{4i N_f N_c I_1(\tilde{m})-2i N_f N_c(k^2-4\tilde{m}^2)I_2(k^2;\tilde{m},\tilde{m})\right\} 
\nonumber\\
& \equiv & Z^2\Pi^r_{\sigma}(k)~,
 \eea
where $\Pi^r_{\sigma}(k)$ denotes the renormalized polarization
function; $I_1(\tilde{m})$ is given by Eq.~(\ref{Im}), and the
integral $I_2(k^2;\tilde{m}, \tilde{m})$ takes the form of Eq.~(\ref{I2}) with $m_M \rightarrow \tilde{m}$.

Analogously,
\bea 
\label{Ppk} 
\Pi_{\pi}(k)&=&-i N_f N_c\tr_{{D}}\int \frac{d^4p}{(2\pi)^4}i \gamma_5 S(p) i \gamma_5 S(p-k) \nonumber\\
&=& Z^2\left\{4i N_f N_c I_1(\tilde{m})-2i N_f N_c k^2 I_2(k^2;\tilde{m},\tilde{m})\right\} \nonumber\\
& \equiv & Z^2\Pi^r_{\pi}(k)~.
 \eea

Note the relation
\bea 
\label{P-ps}
\Pi^r_{\pi}(k^2=0)=\Pi^r_{\sigma}(k^2=4\tilde{m}^2)=\frac{N_cZ\Sigma_H(\tilde{m})}{2G_r\tilde{m}}~.
 \eea
Using the definition of $g^2_{Mqq}$ given in Eq.~(\ref{g2}) and Eqs.~(\ref{Psk}), (\ref{Ppk}), one obtains
\bea 
\label{g2Mqq} 
g^2_{Mqq}=\left[-2iN_f N_c I_2(m_M^2;\tilde{m},\tilde{m})\right]^{-1}~,
\eea
which explicitly shows the large - $N_c$ behavior $g^2_{Mqq}=\mathscr{O}(1/N_c)$ used before.

Finally, using Eqs.~(\ref{DMk}), (\ref{Ppk}), (\ref{P-ps}) and the notation 
$m'^2_M=\{4\tilde{m}^2,0\}$ for $M=(\sigma, \pi)$ we find
\bea 
\label{iZD}
iZ^{-2}{D}_M^{-1}(k)&=&
\left\{-\frac{N_c}{2G_r}+4iN_f N_c I_1(\tilde{m})-2iN_f N_c I_2(k^2;\tilde{m},\tilde{m})(k^2-m'^2_M)\right\} \nonumber\\
&=&\left\{-\frac{N_c}{2G_r}\left(1-\frac{Z\Sigma_H}{\tilde{m}}\right)
-2iN_f N_c I_2(k^2;\tilde{m},\tilde{m})(k^2-m'^2_M)\right\} 
\nonumber\\
&=&
g^{-2}_{Mqq}(k)\left\{k^2-\left[m'^2_M
+\frac{N_c}{2G_r}\left(1-\frac{Z\Sigma_H}{\tilde{m}}\right)g^2_{Mqq}(k)\right]\right\}~,
 \eea
where $g^2_{Mqq}(k)$ is the running meson-quark coupling given by expression Eq.~(\ref{g2Mqq}) with $m^2_M\rightarrow k^2$. 
Thus, we can rewrite the meson propagator as
\bea 
\label{MesPro} 
{D}_M(k)=Z^{-2}\frac{i g^2_{Mqq}(k)}{k^2-m^2_M(k)}~,
\eea
where the running meson mass $m_M(k)$ is given by
\bea 
\label{MesMas} m^2_M(k)&=& 
m'^2_M+\frac{N_c}{2G_r}\left(1-\frac{Z\Sigma_H}{\tilde{m}}\right)g^2_{Mqq}(k) \nonumber\\
&=&
m'^2_M+\frac{N_c}{2G_r}g^2_{Mqq}(k)\left(\frac{m_{0r}}{\tilde{m}}
-3i g^2_{\pi qq}I_2(\tilde{m}^2;\tilde{m},m_{\pi})
+ig^2_{\sigma qq}I_2(\tilde{m}^2;\tilde{m},m_{\sigma})\right).
 \eea
In order to obtain the second line in Eq.~(\ref{MesMas}), we have used Eqs.~(\ref{Hm}), (\ref{gapM}) to get
\bea 
\label{MesMas-2} 
1-\frac{Z\Sigma_H}{\tilde{m}}&=& 1-i8N_f G_r I_1(\tilde{m}) \nonumber\\
&=& \frac{m_{0r}}{\tilde{m}}-3ig^2_{\pi qq}I_2(\tilde{m}^2;\tilde{m},m_{\pi})
+ig^2_{\sigma qq}I_2(\tilde{m}^2;\tilde{m},m_{\sigma})~.
\eea
Obviously, the pole mass of the propagator (\ref{MesPro}) is then
given by the solution of the equation
\bea 
\label{k2Sol} 
k^2=m^2_M(k)|_{k^2=m^2_M}\equiv m^2_M~.
\eea
Thus, in the pole approximation one indeed arrives at the expression (\ref{DMk-ap}).
Finally, Eqs.~(\ref{MesMas}) and (\ref{k2Sol}) give for the meson masses
\bea 
\label{MesMP} 
m^2_{\pi}=\frac{N_c}{2G_r}g^2_{\pi qq}\frac{m_{0r}}{\tilde{m}}+\delta m^2_{\pi}~,
\eea
\bea 
\label{MesMS}
m^2_{\sigma}=4\tilde{m}^2+\frac{N_c}{2G_r}g^2_{\sigma qq}\frac{m_{0r}}{\tilde{m}}+\delta m^2_{\sigma}~,
\eea
where
\bea 
\label{DeltaM} 
\delta m^2_{\pi}&=&-3i\frac{N_c}{2G_r}g^4_{\pi qq}I_2(\tilde{m}^2;\tilde{m},m^2_{\pi})
+i\frac{N_c}{2G_r}g^2_{\pi qq} g^2_{\sigma qq} I_2(\tilde{m}^2;\tilde{m},m^2_{\sigma})~,
\eea
and the expression $\delta m^2_{\sigma}$ follows from Eq.~(\ref{DeltaM}) by replacing factors 
$g^4_{\pi qq}\rightarrow g^2_{\sigma qq} g^2_{\pi qq}$, 
$g^2_{\pi qq} g^2_{\sigma qq} \rightarrow g^4_{\sigma qq}$.

Note that in the chiral limit $m_{0r}=0$, Eq.~(\ref{MesMP}) leads to a non-vanishing pion mass term $m^2_{\pi}=\delta m^2_{\pi}=\mathscr{O}(1/N_c)\neq 0$ 
arising from the meson exchange contributions in the quark gap equation 
(compare Eqs.~(\ref{gapM}), (\ref{MesMas-2})). 
Obviously, the partial cancellation of the $\pi-$ and $\sigma-$ corrections to $\delta m_\pi^2$ is related to the corresponding cancellation in the quark self-energy.

Thus, considering only the dominant quark loop contribution in the polarization function
$\Pi^r_{\pi}(k)$ and taking into account explicit meson exchange contributions in the full quark mass leads to a violation of the Goldstone theorem by a next-to-leading order term $\mathscr{O}(1/N_c)$. 
Clearly, in order to get a final conclusion about $\mathscr{O}(1/N_c)$ corrections to the pion mass, and thus a possible violation of the Goldstone theorem, one should also estimate the contribution by two- and three-loop diagrams of Fig.~\ref{fig:2b}. These correspond to vertex correction and two-meson intermediate state contributions of non-leading order $\mathscr{O}(1)$ to the pion polarization function.

\subsubsection{Two- and three-loop contributions to the pion polarization function
\label{sssec:2-3loop}}

\paragraph{Two-loop diagram with vertex correction
\label{par:2loop}}

Let us first consider the two-loop diagram given by the second term on the r.h.s. of Fig.~\ref{fig:2b} describing a $\mathscr{O}(1/N_c)$ correction to the $\pi qq$  vertex by $(\pi,\sigma)$-meson
exchange. 
For the estimation of a small pion mass we can restrict us to corresponding low-energy considerations. Thus, it is sufficient to use the vertex expression $\Gamma_{\pi qq}(k;p)$ in the low momentum limit, 
$k\rightarrow 0$, $p\rightarrow 0$.
Straightforward calculations give
\bea 
\label{Gpqq} 
\Gamma_{\pi_i qq}(0)=i\gamma_5\tau_i
\left[-g^2_{\pi qq}I_2(0;\tilde{m},m_{\pi})-g^2_{\sigma
qq}I_2(0;\tilde{m},m_{\sigma})\right]~.
\eea
Obviously, we have $\Gamma_{\pi qq}=\mathscr{O}(1/N_c)$.

Inserting this expression into the corresponding diagram of Fig.~\ref{fig:2b}, there remains a standard quark-loop integral as given in Eq.~(\ref{Ppk}) which for $k\rightarrow 0$ yields the following
 $\mathscr{O}(1)$ correction to the renormalized pion polarization function
\bea 
\label{pionPF}
\delta\Pi^r_{1,\pi}(0)=\left[b^{(\pi)}\gamma^{(\pi)}_{\pi qq}+b^{(\sigma)}\gamma^{(\sigma)}_{\pi qq}
 \right]\Pi^r_{\pi}(0)~.
\eea
Here $b^{(\pi,\sigma)}$ are pure numerical weight factors of order $\mathscr{O}(1)$ to be taken from the 
SD equation for $\Pi_{\pi}$, and $\gamma^{(\pi,\sigma)}_{\pi qq}$ denote the $\pi qq$ vertex
corrections due to $\pi$-, $\sigma$-meson exchange given by the bracket of (\ref{Gpqq}).

\paragraph{Three-loop diagram with two meson intermediate states
\label{par:3loop}}

Finally, let us consider the third diagram on the r.h.s. of Fig.~\ref{fig:2b} with a $\sigma\pi$-intermediate state. 
The required $\Gamma_{\pi\pi\sigma}$ vertex has been calculated in Ref.~\cite{2} in the limit of vanishing external momenta and is given by\footnote{In the loop calculation of Ref.~\cite{2}, the quark mass
without meson corrections was denoted by $m$, and $N_c=3$, $N_f=2$. 
Note that $\Gamma_{\pi\pi\sigma}(0)$ is related to the coupling constant $g_{\pi\pi\sigma}$ of 
Ref.~\cite{2} by $g_{\pi\pi\sigma}=g^2_{\pi qq}g_{\sigma qq}\Gamma_{\pi\pi\sigma}$.}
\bea 
\label{Gpps0} 
\Gamma_{\pi\pi\sigma}(0)=\tilde{m}\left[-i4N_f N_c I_2(0;\tilde{m},\tilde{m}) \right]~.
\eea
Using this low-momentum approximation for the vertex, the corresponding next-leading order 
$\mathscr{O}(1)$ contribution to $\Pi^r_{\pi}(0)$ is given by
\bea 
\label{DeltaPr2p} 
\delta\Pi^r_{2,\pi}(0)=-a\Gamma^2_{\pi\pi\sigma}(0)g^2_{\pi qq}g^2_{\sigma qq}I_2(0;m_{\sigma},m_{\pi})~,
\eea
where $a$ is a pure numerical weight factor of order $\mathscr{O}(1)$ to be taken from the diagram. 
Taking the $\mathscr{O}(1)$ corrections from the considered higher-loop diagrams given by 
Eqs.~(\ref{pionPF}), (\ref{DeltaPr2p}) together, we have
\bea 
\label{DeltaPrp}
\delta\Pi^r_{\pi}(0)&=&\delta\Pi^r_{1,\pi}(0)+\delta\Pi^r_{2,\pi}(0) 
\nonumber\\
&=&\left\{-b^{(\pi)}g^2_{\pi qq}I_2(0;\tilde{m},m_{\pi})
-b^{(\sigma)}g^2_{\sigma qq}I_2(0;\tilde{m},m_{\sigma}) \right\}\frac{N_c}{2G_r}\frac{Z\Sigma_H}{\tilde{m}}\nonumber\\
 &&-a\Gamma^2_{\pi\pi\sigma}(0)g^2_{\pi qq}g^2_{\sigma qq} I_2(0;m_{\sigma},m_{\pi})~.
\eea
Adding now $\delta\Pi^r_{\pi}(0)$ to the dominant quark loop contribution $\Pi^r_{\pi}(0)$ in 
Eq.~(\ref{P-ps}) and repeating the steps from Eq.~(\ref{iZD}) to Eq.~(\ref{DeltaM}) leads to the full
pion mass given by
\bea 
\label{M2pi} 
m^2_{\pi}=\frac{N_c}{2G_r}g^2_{\pi qq} \frac{m_{0r}}{\tilde{m}}+\delta m^2_{\pi}+\hat{\delta} m^2_{\pi}~.
\eea

Here the quark-loop contribution $\delta m^2_{\pi}$ is given by Eq.~(\ref{DeltaM}) and related to the meson exchange contributions to the full quark mass $\tilde{m}$, as given in Eq.~(\ref{gapM})
(compare with the second and third terms in the square bracket).
Moreover,
\bea 
\label{DeltaMp2} 
\hat{\delta} m^2_{\pi}=g^2_{\pi qq} \delta\Pi^r_{\pi}(0)~,
\eea
with $\delta\Pi^r_{\pi}(0)$ given by Eq.~(\ref{DeltaPrp}), is related to the two next-to-leading order diagrams of the pion polarization function shown in Fig.~\ref{fig:2b}.

Let us again consider the chiral limit $m_{0r}\rightarrow 0$. 
With the best of our will, we cannot recognize how the two complicated
$\mathscr{O}(1/N_c)$ contributions $\delta m^2_{\pi}$, $\hat{\delta} m^2_{\pi}$ arising from the quark loop diagram and the remaining other two- and three-loop diagrams in Fig.~\ref{fig:2b} should
cancel each other. 
By this reason, we rather come to the conclusion that $m^2_{\pi}=\mathscr{O}(1/N_c)\neq 0$ and that the
Goldstone theorem turns out to be violated.\footnote{It is worth
mentioning that in the interesting perturbative $1/N_c$- approach
of Refs.~\cite{6,10}, using RPA meson propagators and Hartree-type quark
propagators, there occurs a remarkable cancellation between
meson-corrections to the tadpole-like term $\Sigma_H$ in the quark
self-energy and mesonic correction terms to the pion polarization function. 
Since a pure $1/N_c$-correction in the quark mass term (without coupling constant factor $G$), 
as that shown in Eq.~(\ref{gapM}) and leading to $\delta m^2_{\pi}$ of  Eq.~(\ref{DeltaM}) was discarded, 
one found $m^2_{\pi}=0$.}

In fact, a breakdown of the Goldstone theorem in the considered
extended quark model seems to be not so surprising, since a really
self-consistent treatment, based on the chiral Ward identity, would
require equal interaction kernels in both the SD and Bethe-Salpeter (BS) equations. 
This is evidently not the case in the considered NJL model.

Note also that there exist other systematic approaches,
consistently resumming non-perturbative effects in Quantum Field
Theory, which lead to an effective action whose loopwise
expansion introduces residual violations of possible global
symmetries. The latter then give rise to massive would-be
Goldstone bosons \cite{16}.

Finally, it is worth remarking that there have recently been proposed symmetry-improved loopwise expansion techniques \cite{17} for encoding global symmetries. 
In such an approach the extremal solutions for propagators to a loopwise truncated effective
action are then subject to additional constraints, given by the Ward identities due to global symmetries.

\section{Summary and concluding remarks
\label{sec:sum}}
In the present paper, we have investigated the problem of including back-reactions of composite 
mesons in the NJL model. 
For this aim, we first formulated a variational path-integral treatment for deriving an effective action
$\mathscr{A}_{\rm eff}[{D}_{\sigma,\pi};S]$, with ${D}_{\sigma,\pi}$ and $S$ being RPA-like meson and 
full quark propagators. 
In particular, we considered contributions to $\mathscr{A}_{\rm eff}$ given by $2P-$ irreducible loop diagrams with one, two and three meson propagators (compare Fig.~\ref{fig:1} (b) - (e)). 
Using ${D}_{\sigma,\pi}=\mathscr{O}(1/N_c)$, these diagrams contribute then in the spirit of the 
$(1/N_c)$-expansion to $\mathscr{O}(1)$ (Fig.~\ref{fig:1} (b)) and to 
$\mathscr{O}(1/N_c)$ (Fig.~\ref{fig:1} ({c}) - (e)), respectively. 
Subsequently, we have shown that the stationarity conditions 
${\delta \mathscr{A}_{\rm eff}}/{\delta S}=0$,
${\delta \mathscr{A}_{\rm eff}}/{\delta {D}_{\sigma,\pi}}=0$
provide us with coupled SD equations for the mesonic part
$\Sigma_{M}$ of the dynamical quark self-energy and the meson polarization functions 
$\Pi_{\sigma,\pi}$ (compare Figs.~\ref{fig:2a}, \ref{fig:2b}).

It is remarkable that the given variational path-integral
treatment reproduces and extends results of the
$\Phi$-derivable approach \cite{11,12,13,14,15} and, in addition, 
provides a functional foundation for the diagrammatic resummation of
$1/N_c$-corrections \cite{5,6,10} in the NJL model. 
Using the pole approximation for the full quark propagator and the RPA-like meson
propagators containing full quark propagators, we then have
given a low-momentum estimate of the full meson polarization function $\Pi_\pi$ for the pion. 
This required to estimate the one-, two- and three-loop diagrams of order $\mathscr{O}(N_c)$ 
and $\mathscr{O}(1)$ shown in Fig.~\ref{fig:2b}. 
Our results indicate that the coupled SD equations for the quark self-energy and the pion polarization function do not lead in the chiral limit to a vanishing pion mass so that the
Goldstone theorem is violated. 
In fact, this does not seem to be so surprising, if one recognizes that a strictly
selfconsistent approach with use of the chiral Ward identity
should include equal interaction kernels in both the SD- and
Bethe-Salpeter equations. 
We hope to discuss this complicated problem elsewhere.

Finally, it is worth mentioning that the considered variational path-integral approach can potentially be applied to study phase transitions in QCD at finite temperature and density 
(in particular, when hadron dissociation in hot, dense matter is addressed, see \cite{18}) 
or in planar QFT$_3$ of 
graphene-like Gross-Neveu models (see, e.g., Ref.~\cite{19}).

\section*{Acknowledgements}
The authors acknowledge support by the DAAD partnership program between the Humboldt University Berlin and the University of Wroclaw and the hospitality that was extended to them at these Institutions.
We thank N.T. Gevorgyan for help in preparing the manuscript.
D.B. was supported in part by the Polish NCN under contract number DEC-2011/02/A/ST2/00306
and by the MEPhI Academic Excellence Project under contract number 02.a03.21.0005.

\end{document}